# Stock price direction prediction by directly using prices data: an empirical study on the KOSPI and HSI

Yanshan Wang

School of Industrial Management Engineering,
Korea University,
Seoul, 136-713, Korea
E-mail: yansh.wang@gmail.com

**Abstract:** The prediction of a stock market direction may serve as an early recommendation system for short-term investors and as an early financial distress warning system for long-term shareholders. Many stock prediction studies focus on using macroeconomic indicators, such as CPI and GDP, to train the prediction model. However, daily data of the macroeconomic indicators are almost impossible to obtain. Thus, those methods are difficult to be employed in practice. In this paper, we propose a method that directly uses prices data to predict market index direction and stock price direction. An extensive empirical study of the proposed method is presented on the Korean Composite Stock Price Index (KOSPI) and Hang Seng Index (HSI), as well as the individual constituents included in the indices. The experimental results show notably high hit ratios in predicting the movements of the individual constituents in the KOSPI and HIS.

**Biographical notes:** Yanshan Wang received his BE degree in Computer Science from Harbin Institute of Technology and a ME degree in Management Engineering from Korea University, in 2010 and 2012, respectively. He is currently working towards his PhD degree in the Department of Management Engineering at the Korea University. His research interests include financial engineering, nonlinear programming, information retrieval and machine learning.

## 1 Introduction

Nowadays, investors have been more keenly aware of risk than any time in the past due to the non-stationary and chaotic stock markets under the impact of the US subprime crisis. Simultaneously, they hope to gain a great profit from the investments. But, it is extraordinarily difficult to perform better than skilled and knowledgeable competitors in a stock market. Better *stock prices direction prediction* is a key reference for better trading strategy and decision-making by ordinary investors and financial experts (Kao et al., 2013). Apart from the *stock price direction prediction*, the *stock market index direction prediction* is regarded as one of the crucial issues in recent financial analysis studies. The *stock market index measures* overall market behaviour through selected stocks representing the market. Accurate index direction prediction provides investors with information regarding the expectations about the movement behaviour of the economy and return obtained by a specific investing portfolio (Pathak et al., 2011). Also it is an early warning system for investors, notably for short-term investors, against sudden drops in the market.

The emergence of machine learning and artificial intelligence algorithms has made it possible to tackle computationally demanding mathematical models in *stock price direction prediction*. Frequently adopted methods include artificial neural networks (ANNs), Bayesian networks, and support vector machine (SVM). Amongst them, ANNs have drawn significant interests from several researchers in the stock price forecasting in the past decades. The ANNs are robust in model specification compared to parametric models, which makes it frequently applied in forecasting stock prices and financial derivatives. Guresen et al. (2011) reported the validity of ANNs in stock market index prediction. Cheng et al. (1996) forecasted US treasury bond with a ANNs-based system. Grudnitski and Osburn (1993) applied ANNs to predict gold futures prices. The drawback of the price prediction is that the price is highly volatile so as to result in large regression errors. Compared to the price prediction, the *stock direction prediction* is less complex and more accurate (Ou and Wang, 2009). The stock direction prediction has been recently addressed in several research articles, which consider different variants of ANNs (Saad et al., 1998). However, one drawback of ANNs is that the efficiency of predicting unexplored samples decreases rapidly when the neural network model is too over-fitted by available observations. In other words, the noisy stock information may lead ANNs to a complex model, which might result in the over-fitting problem.

The predominant methods in the stock market direction prediction are the approaches based on SVM (Huang et al., 2005; Kim, 2003; Tay and Cao, 2001). Since the SVM implements the structural risk minimisation principle, it often achieves better generalisation performance and lower risk of overfitting than the ANNs (Cortes and Vapnik, 1995). This point in regard to stock prediction is well braced by Kim (2003). His experiment showed that the SVM outperformed the ANNs in predicting future direction of a stock market and yet reported that the best prediction performance that he could obtain with SVM was 57.8% in the experiment with the Korean composite stock price index 200 (KOSPI 200). Huang et al. (2005) reported remarkable performance of 75% hit ratio by using a SVM-based model to predict Nihon Keizai Shimbun Index 225 (NIKKEI 225) in a single period. However, the shortcoming in the most of literature is that the testing used was conducted within the in-sample datasets, or that the out-of-sample testing was on small data sets which were unlikely to represent the full range of market behaviours. Thus it is difficult to assess the average performances of their models in multiple periods such as Kim nor Huang et al. conducted the experiment in that setting. In order to refrain from limited sample selection, our experiments computed the one-day-ahead predictions using rolling windows of data to ensure that the predictions are made using all the information available at that time, while not incorporating old data that are probably no longer relevant in the context of a dynamic, rapidly evolving stock market.

A major drawback of SVM for the direction prediction is that the input variables lie in a high-dimensional feature space, ranging from hundreds to thousands. The storage of the variables requires a lot of memory and computation time. Specifically, a stock market consists of several hundreds of stocks, which leads to the high dimensionality of the variables. Therefore, it is of considerable importance to conduct dimension reduction to acquire an efficient and discriminative representation before classification. Under the dimensionality reduction, *curse of dimensionality* could be effectively managed (Cortes and Vapnik, 1995). A common unsupervised feature extraction method is principal component analysis (PCA) (Pearson, 1901) by which principal components are obtained through the manipulation of original data. The PCA has been widely used to deal with high dimensional data sets in many areas, such as protein dynamics reduction, spectral data reduction, and face patterns reduction. Interestingly, the adaptation of the PCA feature selection to stock prices data analysis is rarely found, to the best of our knowledge.

In stock prices data, there exits a common phenomenon that is called *co-movement* between stocks due to the institutional investors¡¯ common ownership of subsets of stocks in their portfolios (Pindyck and Rotemberg, 1993). Shiller (1989) showed the *co-movements* of returns between the USA and UK markets using simple regression tests. The same phenomenon may also exist between the US stock market the Asian markets, as reported by Liu et al. (1998). However, these methods are not explicit to verify the *co-movements* and find the *co-moved* stocks. The *co-movement* implies that the utilisation of PCA is essential for finding *co-moved* stocks among highly correlated stocks. We are the first to verify the *co-movement* phenomenon by showing the principal components, to the best of our knowledge.

As a matter of fact, the *co-movement* exits not only between stocks in a domestic market (internal) but also between two tightly connected stock markets (external). This facts stimulate us to consider both internal and external factors for predicting individual stocks and market index directions. The macroeconomic indicators [such as consumer price index (CPI), gross national product (GNP) and gross domestic product (GDP)] may be high internal impact factors for the prediction. However, daily data of those macroeconomic factors are impossible to obtain and analyse in reality. For simplicity and generality, we only handle stock prices data which are timely and easy to access. As for the external factors, we take daily S&P 500 index values and exchange rates (EXR) into account. Both factors can be obtained and managed easily. Thus, the method in this article secondly contributes to the stock prediction in practical aspect compared with most of the state-of-the-art approaches.

This paper is organised as follows. In the next section, we provide a brief overview of PCA and SVM and describe how they are integrated in our model. In Section 3, we present detail descriptions on the empirical experiment, which includes the detailed design of experimental data and experiment results. The last section concludes this paper with some discussions.

## 2 An integrated model

The structure of the proposed model is shown below in Figure 1. Let $x_i \in \mathbb{R}^p$ denote a column vector of the daily rates of return of stock $i$, $i = 1, \ldots, n$, which is obtained from $p$ daily market observations. The matrix $X = (x_i)^T$ can be reduced to the principal component matrix $Y = (y_k)^T$, $k = 1, \ldots, m$, $m \ll n$ by minimising the variance of the linear transformation of $X$. Define the contribution rate of the $k^{th}$ principal component as $\frac{\lambda_k}{\sum_{i=1}^{n} \lambda_i}$ where $\lambda_k$ represents the variance of $y_k$. The cumulative rate of the first $m$ principal components is $\frac{\sum_{i=1}^{m} \lambda_i}{\sum_{i=1}^{n} \lambda_i}$.

Along with these principal components, internal factors and external factors $\mathbf{F} = (F_1, F_2, \ldots, F_h)^T$ are utilised as input data, i.e., $\{\mathbf{Y}, \mathbf{F}\}$. Considering the *co-movement* property in a market, we find that the *co-moved* stocks are informative as internal factors. Besides, since the market index itself is a beacon of the domestic economy and tread, it is also informative for forecasting. In addition to the internal factors, external factors also play important role in the Asia stock markets. Here we consider two foremost economic phenomena. The USA is the largest cooperative partner for Asian countries, including Korea and Hong Kong, thus the conditions of the US financial market have significant impacts on Asian stock market. The other is the EXR that also has strong influence on the imports and exports of products. The changes of trading relations in turn affect the domestic stock markets. Thus, the external factors used in our study include the S&P 500 index in the US stock market, the best representation of the US economy, and the exchange rate, the symbol of trades between countries. Details of the data are explained in Section 3.

The input to SVM is the data set $\mathbf{R} = (\mathbf{r}_j, w_j)$, where $\mathbf{r}_j = (\mathbf{Y}_j, \mathbf{F}j)$ ($j = 1, \ldots, p$) is a row vector denoting $j^{th}$ daily data in $p$ observation days. $w_j \in 0,1$ is a binary variable that represents the upward or downward direction of the stock market movement of the $j^{th}$ day. The downward direction is represented by 0 and the upward by 1. The input data is carefully divided into two parts, i.e., training data and testing data. As addressed by financial analysis recently, the data periods in most computer technique related articles are selected limitedly. In order to refrain from limited sample selection, the training data and testing data are goes parallel using rolling windows of ten years data to ensure that the predictions are made using all the information available at that time. Besides, unlike several studies testing in-sample data, we compute the one-day-ahead predictions, i.e., out-sample data. The details of data periods and how they are divided into training and testing data are explained in the next section.

The training data is utilised to acquire a classifier by training SVM. The classifier function of stock movement directions is defined as

$$\text{dir} = f(\mathbf{r}) = \text{sgn}\left(\sum_{j=1}^{p} w_j \alpha_j^* \mathbf{r}_j^T + v^*\right), \tag{1}$$

**Figure 1** Procedure of the proposed PCA-SVM integrated model

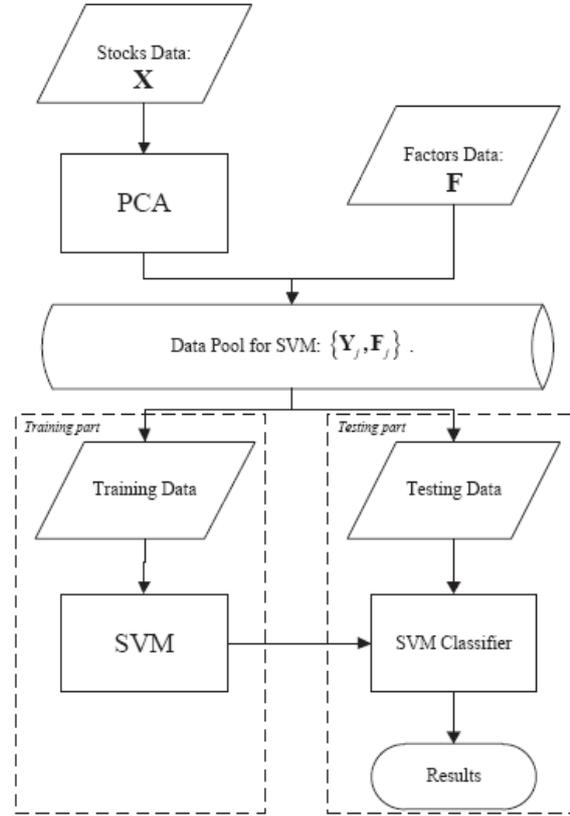

where $\alpha^*_i$'s and $v^*_i$'s are optimal values of Lagrange multipliers and intercepts of the corresponding hyperplanes, respectively. By introducing kernel tricks, the nonlinear decision function for a stock direction prediction becomes

$$\text{dir} = f(\mathbf{r}) = \text{sgn}\left(\sum_{j=1}^{p} w_j \alpha^*_j \mathbf{K}(\mathbf{r}_j, \mathbf{r}) + v^*\right). \quad (2)$$

The selection of kernel function is addressed in Section 3.

The testing data is used to test the model according to the classifier $f(\mathbf{r})$. In reality, the training model can be designed to update real-timely so as to make full use of the present information.

## 3 Empirical experiment

In this section, we present the empirical stock price data sets, time periods and data designs for the proposed method. In order to avoid the fact that some markets are less efficient than others, we test empirical experiments on two representative Asia stock markets: the Korean stock market and the Hong Kong stock market. The experiments aim to forecast the directions of daily movements of the stock price indices and of individual stocks.

## 3.1 Experiment data

### 3.1.1 Stock market indices

The Korea composite stock prices index 200 (KOSPI 200) and the Hang Seng index (HSI) are utilised since they represent the overall performance of the Korean and the Hong Kong stock market. As an underlying index for stock index futures and options, KOSPI 200 consists of the 200 companies chosen from all stocks in the KRX-Stock Market. The KOSPI 200 represents a broad cross-section of Korean industries and provides an effective means for investors to avoid potential market risks. Thus, the 200 individual stocks¡¯ daily rates of return are utilised. Similarly, the HSI is the main indicator of the overall market performance in Hong Kong and the 48 constituents are processed in our empirical approach. We note that all the data are collected from publicly available sources in the internet. The KOSPI 200 data set are collected from KRX Korea Exchange (http://eng.krx.co.kr). HIS data set are collected from YAHOO! Finance (http://finance.yahoo.com).

### 3.1.2 Factors

As addressed in the preceding and the structure of the method (Figure 1), internal factors *co-movement* stocks considered amongst the stock market and the market index. For forecasting indices, we use the lagged daily prices for indices and the overall constituents. As for individual stock prediction, we use the lagged daily prices for the target-excluded constituents besides indices. The external factors used in our study include the S&P 500 index in the US stock market and the exchange rate of US dollars to Korean Won (USDKRW) or Hong Kong Dollar (USDHKD). The S&P 500 index data and USDKRW (USDHKD) data are downloaded from YAHOO! Finance (http://finance.yahoo.com) and International Monetary Fund (http://www.imf.org), respectively. Because the opening days of the markets in Asia markets and the USA are different, the data are aligned based on the Asia markets¡¯ timeline. The dealing strategy is that redundant daily data are deleted while missing daily data are filled by the previous closing price.

### 3.1.3 Time periods

In order to avoid limited sample selection, the data sets tested in our empirical experiment are gathered for the time periods of 2002 to 2011. Unlike most of the studies that tested machine learning methods within in-sample data, we tested the proposed method using out-sample data. Besides, a rolling window design of data was used so as to fully check the performance of the method and ensure that the predictions were made using all the information available at that time. The time periods of the rolling window data are shown in Table 1. In short, we treat three years¡¯ training data and the following year¡¯s testing data as a window which slides from the first year until the end of the ten years period.

*3.1.4 Data pre-process*

Instead of using the daily rate of return directly, we transform it into an n-day relative difference in percentage of price (RDP). The advantage of the transformation is that the distribution of the transformed data will be more symmetrical and close to a normal distribution (Tay and Cao, 2001).

In this paper, the RDP values are determined based on three-day-lagged (RDP-3) values for the indices and EXR, and one-day-lagged (RDP-1) for the constituents of KOSPI and HSI. The reason of choosing RDP-3 values for the formers is that market indices and EXR changes always have delayed-effects on the index values (Shiller, 1989). Since the constituents serve as market comprising elements, the *co-movements* between the elements affect the market itself immediately. Therefore, a shorter lagged period is selected. The direction to forecast is the sign of one-day-ahead RDP, which is denoted as RDP+1. The detailed calculations for all the indicators are given in Table 2.

**Table 1**  Corresponding time period of training and testing data

| Round | Training period | Testing period |
| --- | --- | --- |
| 1 | 2002-1-1 2005-1-1 | 2005-1-1 2006-1-1 |
| 2 | 2003-1-1 2006-1-1 | 2006-1-1 2007-1-1 |
| 3 | 2004-1-1 2007-1-1 | 2007-1-1 2008-1-1 |
| 4 | 2005-1-1 2008-1-1 | 2008-1-1 2009-1-1 |
| 5 | 2006-1-1 2009-1-1 | 2009-1-1 2010-1-1 |
| 6 | 2007-1-1 2010-1-1 | 2010-1-1 2011-1-1 |
| 7 | 2008-1-1 2011-1-1 | 2011-1-1 2012-1-1 |

**Table 2**  Input and output indicators

| Indicator | Calculation |
| --- | --- |
| *Input indicators:* | |
| RDP-1 | $\dfrac{\mathbf{r}_j - \mathbf{r}_{j-1}}{\mathbf{r}_{j-1}} \times 100\%$ |
| RDP-3 | $\dfrac{\mathbf{r}_j - \mathbf{r}_{j-3}}{\mathbf{r}_{j-3}} \times 100\%$ |
| *Output indicators:* | |
| RDP+1 | $\dfrac{\mathbf{r}_{j+1} - \mathbf{r}_j}{\mathbf{r}_j} \times 100\%$ |

## 3.2 Direction forecasting for indices and individual stocks

Our task is to forecast the daily direction of KOSPI (or HSI) and the movement directions of the constituent stocks in KOSPI (or HSI). Based on equation (2), the direction function can be written as follows

$$\mathrm{dir}^{RDP+1} = f\left(F_{\mathrm{Index}}^{RDP-3}, F_{\mathrm{S\&P500}}^{RDP-3}, F_{\mathrm{EXR}}^{RDP-3}, Y_1^{RDP-1}, Y_2^{RDP-1}, \ldots Y_k^{RDP-1}\right), \qquad (3)$$

where $F_{\mathrm{Index}}^{RDP-3}$, $F_{\mathrm{S\&P500}}^{RDP-3}$, $F_{\mathrm{EXR}}^{RDP-3}$ are the RDF-3 values of the KOSPI (or HSI), the S&P 500 Index and the exchange rates of USD against KRW (or HKD), respectively. $Y_1^{RDP-1}, Y_2^{RDP-1}, \ldots, Y_k^{RDP-1}$ are principal components of all the constituents with the input defined by RDF-1 value of the stocks in KOSPI (HSI). $\mathrm{dir}^{RDP+1}$ is a categorical variable that is obtained from the SVM classifier and represents the movement directions of the prices, i.e.

$$\mathrm{dir}^{RDP+1} = \begin{cases} 1, & \text{price increases,} \\ 0, & \text{price decreases,} \end{cases} \qquad (4)$$

Note that the direction is assigned to 1 if the closing value is the same as the previous closing value.

**Figure 2** Contribution rates of principle components for the KOSPI (see online version for colours)

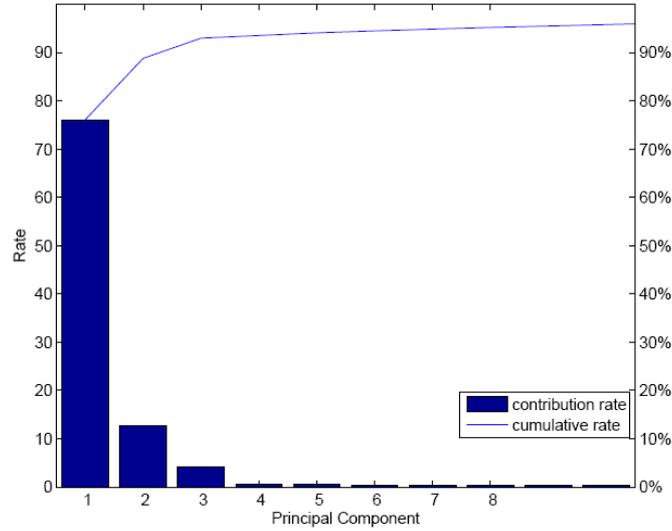

**Figure 3** Contribution rates of principle components for the HIS (see online version for colours)

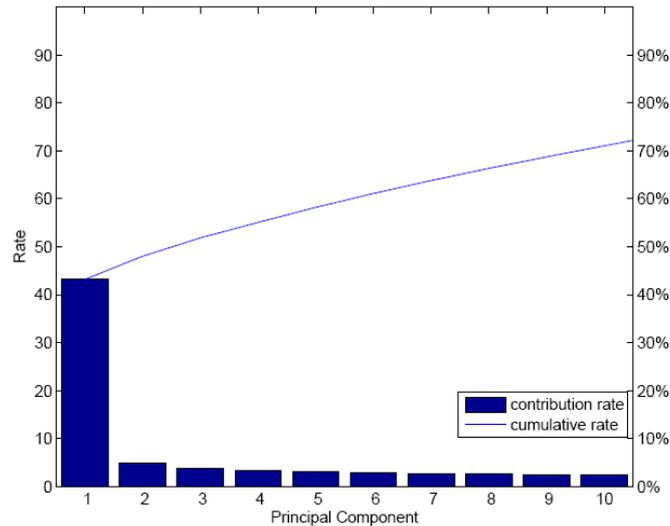

### 3.3 Experiment results

#### 3.3.1 Results of the PCA

Figures 2 and 3 depict the contribution rate (histogram) and cumulative distribution (line) of the first several components of the constituents in KOSPI and HSI over the whole time period respectively. The cumulative contribution is plotted by accumulating the contribution rate. From the following figures we find that the first component has over 70%contribution for the KOSPI and the first ten components have over 70% for the HSI. Thus, the first 1 while the first ten components are chosen as principal components for the KOSPI and HIS to predict the $dir^{RDP+1}$, respectively.

#### 3.3.2 Verification of the co-movement

In this subsection, a two-dimensional illustrative example of the PCA is given for visualising the principal components, as shown in Figure 4. The first two stocks in KOSPI, Samsung Elec. and Hyundai Mtr., are selected under observations over the whole periods. All the data are scaled to have a mean of zero. So, the principal components can be represented by two orthogonal arrows passing through the origin with greatest variance. The thick line represents the first principal component and the dashed line represents the second principal component. Both arrows are the vectors that are derived from the eigenvectors of the covariance matrix that are scaled by the corresponding eigenvalue. Then, the space is rotated so that principal components are aligned with the coordinate axes. In doing so, the data are uncorrelated in the principal component space.

**Figure 4** The plots of the first two principal components in KOSPI: SamsungElec and HyundaiMtr (see online version for colours)

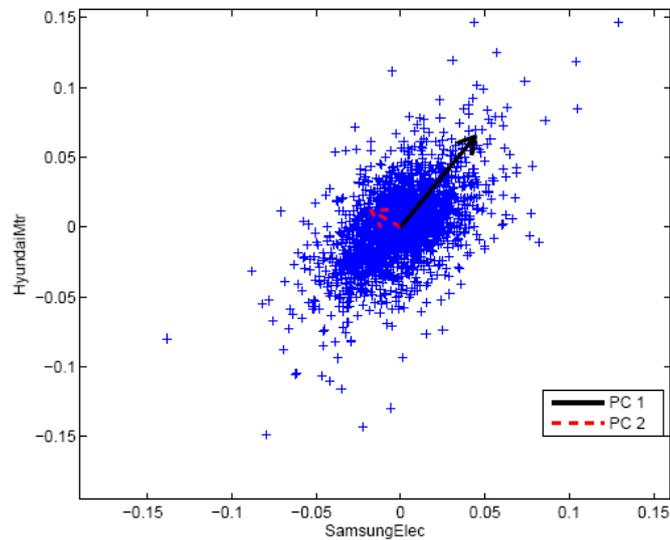

A biplot is helpful in digging some hints of the PCA. The biplot plots the projection of the loadings of the stocks on to the first two principal components. The biplot figure for the KOSPI and HIS are shown in Figures 5 and 6, respectively. The plots reflect the phenomenon of *co-movements*, because the *co-moved* constituent stocks are near to each other and construct to a cluster. Empirically, we observe that those highly correlated stocks in one cluster are from the companies in a similar domain area (e.g., SKTelecom, KT and KT&G) or that some of the correlated companies are sub-companies and sub-branches of the other companies (e.g., Samsung braches and Samsung Co., LG branches and LG Co.).

### 3.3.3 Results of the direction forecasting on the KOSPI and HSI

There are two decisions to make for SVM classifier. One is the choice of a kernel function to use, i.e., linear, polynomial or radial basis function (RBF). The other is the selection of parameter $C$. Several studies have suggested the RBF kernel function (Huang et al., 2005; Kim, 2003; Tay and Cao, 2001) which is also used in our experiment. As for the parameter $C$, a small value causes under-fitting of the training data and a large value causes over-fitting. Therefore, a value between 0.1 and 1,000 is known to be the appropriate choice for the parameter $C$. In our experiment, $C = 100$ is used since it is tested in the previous studies (Kim, 2003; Tay and Cao, 2001).

Given the RBF kernel and $C = 100$ for the SVM classifier, we examine the effectiveness of the proposed method in forecasting indices. Comparisons of the hit ratios are shown in Figure 7 for the KOSPI and HIS. To compare with the performance of the SVM, we also test the ANN and random walk (RW) as the benchmark. Besides, we input identical principal components for both the SVM and ANN to verify the influence of the PCA.

**Figure 5** Biplot of the KOSPI (see online version for colours)

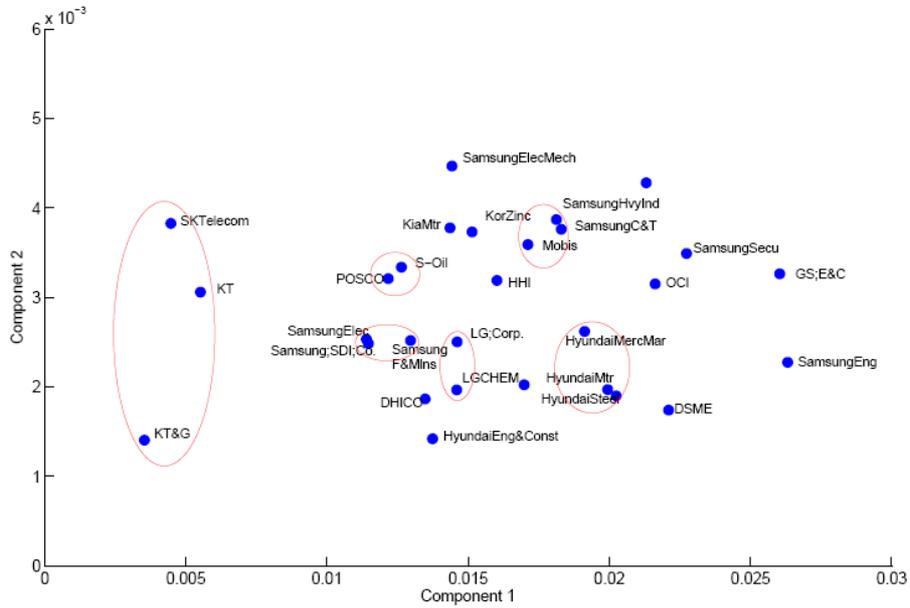

Notes: The horizontal and vertical axes represent the first and the second principal components, respectively. The *co-moved* stocks are circled by the red ellipses.

**Figure 6** Biplot of the HIS (see online version for colours)

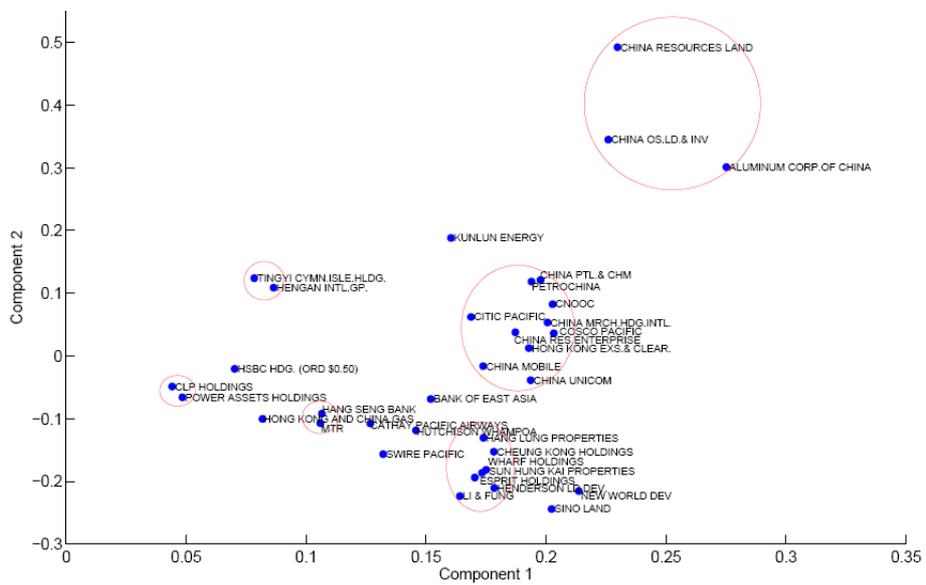

Notes: The horizontal and vertical axes represent the first and the second principal components, respectively. The *co-moved* stocks are circled by the red ellipses.

**Figure 7** Hit ratios of forecasting KOSPI and HIS

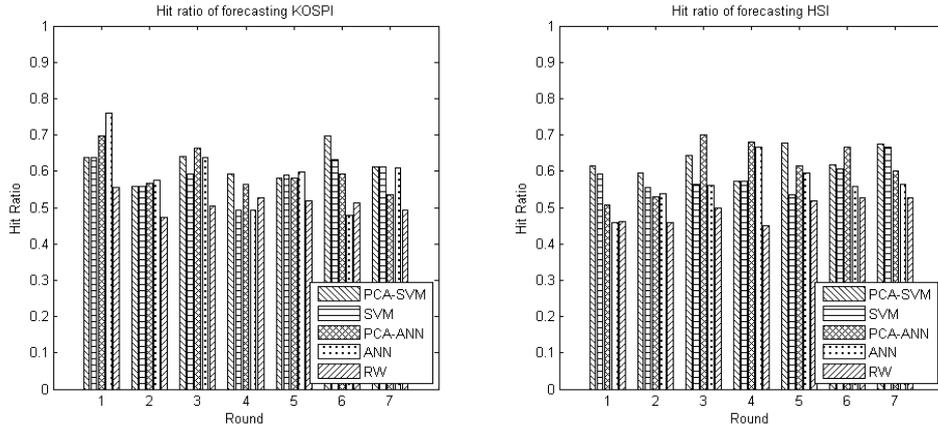

**Figure 8** Average value and standard derivation of the hit ratios

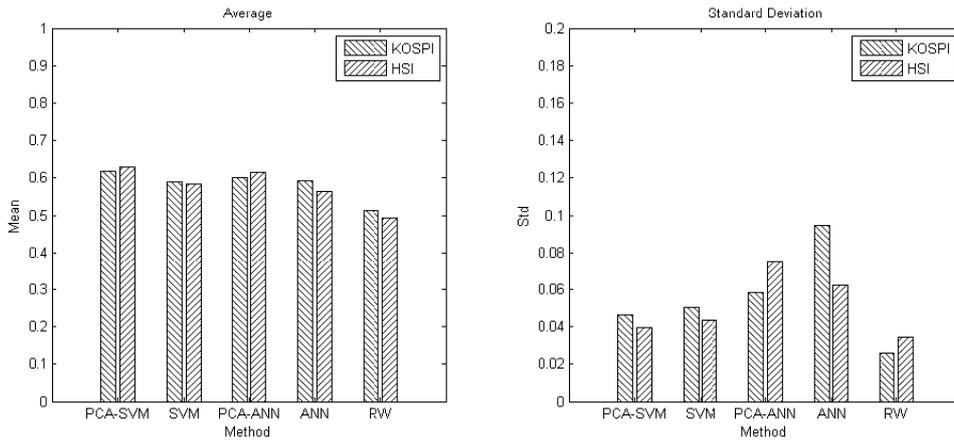

From the tables, the RW model performs moderately. The influence of the PCA is obvious positive since both the PCA-SVM and PCA-ANN outperforms the original SVM and ANN respectively. The average hit ratios of the proposed PCA-SVM on forecasting the KOSPI and HIS are minorly better than the PCA-ANN, as shown in Figure 8. However the derivations of the ANN-based methods are bigger than the SVM. This fact reflects the drawbacks of the ANN that refer to the volatility and over-fitting problems, which is clearer from the tables. On the other hand, we emphasise that the at round 4, the period of 2008 when the subprime crisis occurred, the PAC-SVM performs moderately during that period but still better than the RW model. However, the ANN is untrustworthy because of the enormous difference between two indices forecasting results.

*3.3.4 Results of the direction forecasting on the constituents*

The second experiment of forecasting directions of the constituents of the KOSPI and HSI is carried out with the proposed PCA-SVM, ANN and RW. In this paper, we report four sample constituents from the KOSPI [SamsungElec (SAEL), HyundaiMtr (HYMT)] and the HIS [BANK OF EAST ASIA (BOEA), CITIC PACIFIC (CIPA)] shown in Figures 9 and 11, as well as the average values and standard derivations in Figures 11 and 12. Unlike the market indices direction prediction, the hit ratios of forecasting individual constituents are averagely better. The influence of the PCA is also noticeable as well as that of the previous section. The PCA-SVM gives higher average hit ratios with lower standard derivation compared with the PCA-ANN. All the testing experiments show that the methods using the PCA perform better. To summarise, the proposed method are considerably trustworthy in forecasting movement directions.

**Figure 9** Hit ratios of forecasting SAEL and HYMT in KOSPI

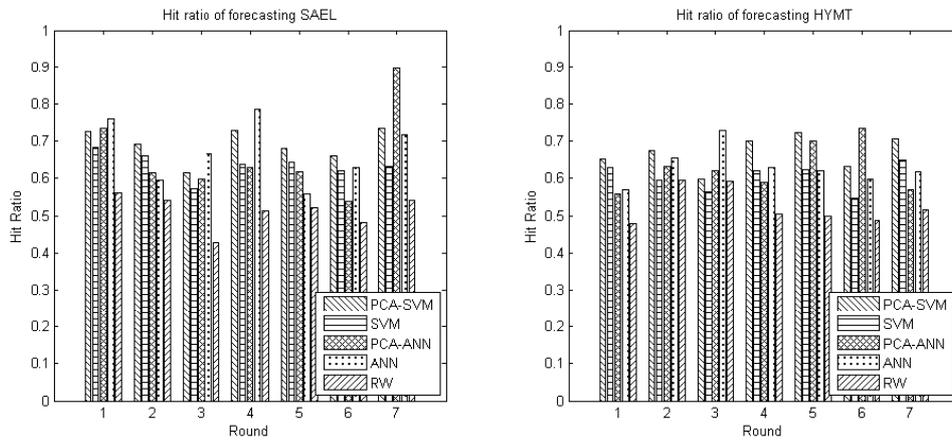

**Figure 10** Average value and standard derivation of the hit ratios

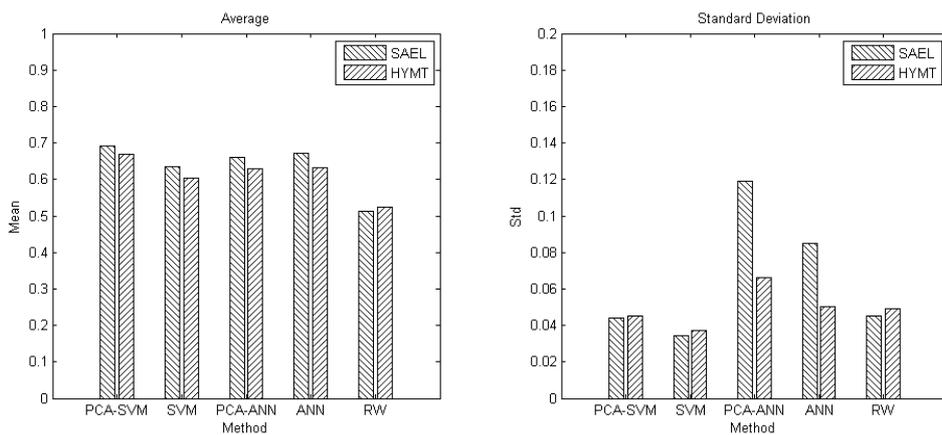

**Figure 11** Hit ratios of forecasting BOEA and CIPA in HIS

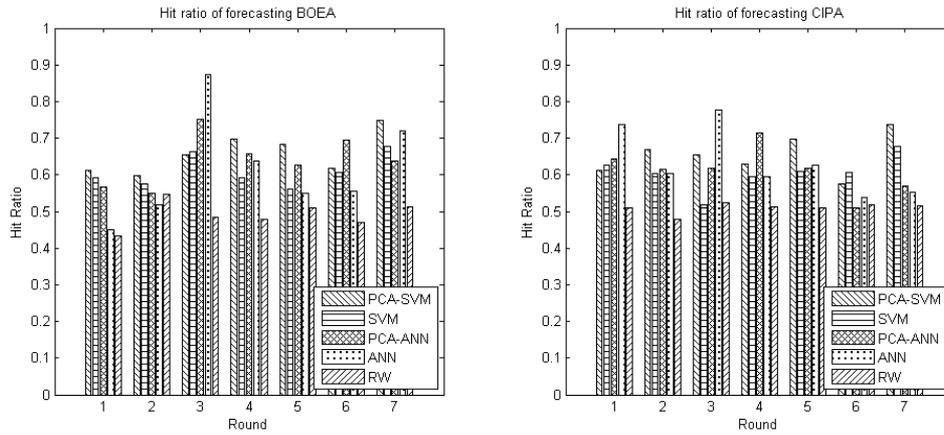

**Figure 12** Average value and standard derivation of the hit ratios

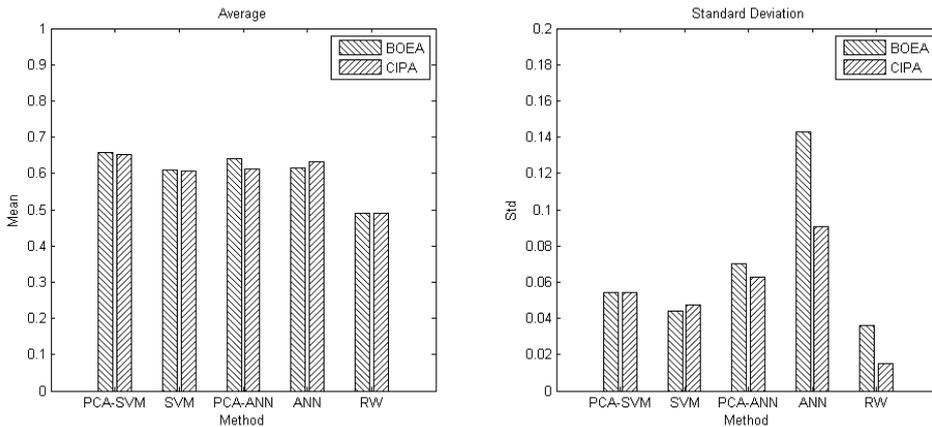

The accuracy of forecasting directions of constituent is significantly higher (at confidence level 0.05) than index, which is distinct from the general understanding that index forecasting is easier. We can account for this observation by analysing the stock price data. Since the constituent stocks prices are more volatile than index values, it is easier for classifiers to find patterns. Thus, the forecasting the constituent stocks gives better accuracy.

## 4 Conclusions and future research

In this paper, we have proposed a PCA-SVM integrated model to forecast the directions of the stock market indices and the individual stock prices. In the model, the principal components identified by the PCA are used along with internal and external financial factors in SVM for forecasting. We have also presented an extensive empirical experiment based on the KOSPI and HSI. The results of the empirical experiments show that the proposed method provides markedly high hit ratios for forecasting movement directions of the constituents in the KOSPI and HSI. Since our experiments computed the one-day-ahead predictions using rolling windows data of a long period, the results are not the product of limited sample selection but reliable with all the available information at that time. Our results also verifies the *co-movement* effect between the Korean (or Hong Kong) stock market and the US stock market because of the usage of S&P 500 and exchange rates.

As a future study, a theoretical study on the performance of the proposed method is of worth. The clustering of the *co-moved* stocks according to the biplot needs a further investigation. The theoretical analysis of the better performance on forecasting the constituents is also worth studying. Moreover, other feature selection methods, for example, deep belief networks (DBN), may be also efficient to extract the features of the stock prices for classifiers, which is subject to another future research.